\documentclass{pasa}%
\usepackage{graphicx}
\usepackage{lscape}
\usepackage{scalefnt}
\title[Orbits of selected globular clusters in the Galactic bulge]{Orbits of selected globular clusters in the Galactic bulge}
\author[P\'erez-Villegas et al.]{ A. P\'erez-Villegas{$^1$}, L. Rossi$^2$, S. Ortolani$^{3,4}$,  S. Casotto$^{3,4}$, B. Barbuy$^1$ and E. Bica$^5$\thanks{This is an example of author footnote}\\
\affil{$^1$Universidade de S\~ao Paulo, IAG, Rua do Mat\~ao 1226,
Cidade Universit\'aria, S\~ao Paulo 05508-900, Brazil}%
\affil{$^{2}$Centre for Astrophysics and Supercomputing, Swinburne University of Technology, Hawthorn, VIC 3122, Australia}
\affil{$^3$Dipartimento di Fisica e Astronomia, Universit\`a di Padova, I-35122 Padova,
 Italy}%
\affil{$^4$INAF-Osservatorio Astronomico di Padova, Vicolo dell'Osservatorio 5,
I-35122 Padova, Italy}%
\affil{$^5$Universidade Federal do Rio Grande do Sul, Departamento de Astronomia,
CP 15051, Porto Alegre 91501-970, Brazil}}%
\jid{PASA}
\doi{10.1017/pas.\the\year.xxx}
\jyear{\the\year}

\begin{document}%
\begin{abstract}
We present orbit analysis for a sample of 8 inner bulge globular clusters, together with one reference halo object. 
We  used proper motion values derived from long time base CCD data.
 Orbits are integrated in both an axisymmetric model and a model including the Galactic bar potential. The inclusion of the bar proved to be essential for the description of the dynamical behavior of the clusters. We use the Monte Carlo scheme to construct the initial conditions for each cluster, taking into account the uncertainties in the kinematical data and distances. The sample clusters show typically maximum height to the Galactic plane below 1.5 kpc, and develop rather eccentric orbits.
Seven of the bulge sample clusters share the orbital properties of the bar/bulge, 
having perigalactic and apogalatic distances, and maximum vertical excursion from the Galactic plane inside the bar region. 
NGC 6540 instead shows a completely different orbital behaviour, having a dynamical signature of the thick-disc.
 Both prograde and prograde-retrograde orbits with respect to the direction of the Galactic rotation were revealed,
which might characterize a chaotic behaviour. 
\end{abstract}
\maketitle%
\section{Introduction}
\label{sec:intro}

The orbital evolution of globular clusters (GCs) in the Galaxy, combined with  kinematics and stellar population analyses, can provide important information to decipher the history of our Galaxy.  The Galactic bulge contains a significant fraction of the globular cluster population ($\sim$25\% of GCs) in the Galaxy (Bica et al. 2016). This fraction may increase significantly in coming years with analyses of recent 84 VVV bulge GC candidates (Minniti et al. 2017a, Minniti et al. 2017b).
 In spite of this, very few attempts have been made so far to study the orbits of bulge clusters, previously prevented mainly due to high reddening, crowding and distance effects. 

Two main pieces of information are needed in order to determine the Galactic orbit of a globular cluster: the Galactic position and velocity components of the cluster and a mass model of the Galaxy. The position and velocity components form a six--dimensional vector, called initial state vector, representing the initial conditions required to solve the equations of motion (EoM) associated with the gravitational potential generated by the host galaxy. In this approach, the cluster is treated as a test particle.  This is the typical approximation to study GCs orbits (Aguilar, Hut \& Ostriker 1988; Gnedin \& Ostriker 1997; Gnedin, Lee \& Otriker 1999; Dinescu et al. 2001,
 Casetti-Disnescu et al. 2007, 2013; Pichardo et al. 2004; Allen et al. 2006, 2008; Ortolani et al. 2011; Moreno et al. 2014). Even in works where the destruction rates and tidal stripping are studied in depth (e.g. Moreno et al. 2014), the calculation of the orbits is carried out under the assumption that the cluster acts as a test particle, and it is over such invariant orbit that individual clusters (including N-body model clusters) are calculated (J{\'{\i}}lkov{\'a} et al. 2012; Martinez-Medina et al. 2018). In general, for massive clusters (i.e. the ones we see today), the effect of mass loss on the orbit is negligible. Even in clusters that go through substantial mass loss, where
part of the mass loss
 is expected to be due to evaporation across all the surface of the cluster and part to tidal effects (bipolar),
neither is expected to affect the center of mass of the remaining cluster, and consequently it will not affect the orbit.

In our previous paper (Rossi et al. 2015, hereafter Paper I) we presented the relative
 proper motions to the bulge field stars of ten globular clusters located in the Galactic bulge, namely Terzan 1, Terzan 2, Terzan 4, Terzan 9, NGC 6522, NGC 6558, NGC 6540, Palomar 6, AL 3 and ESO 456--SC38. 
The clusters AL3 and ESO 456--SC38 are not analysed in the present work
 because they have no radial velocity determinations in the literature. By combining information on the proper motion with position components and radial velocities, we derived the initial state vector of eight of the bulge clusters in our sample. For comparison purposes, we also included in the analysis the inner halo cluster NGC 6652, for which the initial state vector has been computed from proper motion data by Sohn et al. (2015).

 The main goal of this follow--up paper is to construct the orbit of the bulge globular clusters with the purpose of checking whether
 they are trapped by the Galactic bar or not, and at the same time, we seek for correlations between the distribution of clusters and
of their 
kinematic properties with metallicity, that could be interpreted as due to the dynamics produced by the Galactic bar.


The present work is structured as follows. In Section \ref{sec:data}, we present the data available for the sample GCs, 
that were employed in the calculations. In Section \ref{sec:methods} the adopted mass models of the Milky Way and the process followed to integrate the orbits of the clusters and their orbital parameters are described. In Section \ref{sec:cluster_orbits} the orbit of the clusters are presented and their main features are briefly discussed. Section \ref{sec:correlations} includes an analysis of correlations between the orbital and chemical properties of the sample, while in Section \ref{sec:discussion} the results are summarized and discussed.

\section{Data for the Globular Clusters}
\label{sec:data}

Table \ref{tab:data} contains the cluster parameters employed in this work. Equatorial $(\alpha,\delta)$ and Galactic $(l,b)$ coordinates are given in Columns 2 and 3, and Columns 4 and 5, respectively. The proper motions relative to the bulge field stars, $\mu_{\alpha}^*= \mu_{\alpha}\cos \delta$, $\mu_{\delta}$, in Columns 6 and 7, given in Paper I, except for NGC 6652, whose values are taken from Sohn et al. 2015, where the proper motions are absolute. The heliocentric radial velocity $v_r$, in Column 8, is taken from the compilation by Harris (2010). The heliocentric distance and metallicity in Columns 9 and 10 are from Bica et al. (2016,
and references therein).

\begin{table*}
\scalefont{0.85}
\caption{Globular cluster data \textsuperscript{a}.}
\label{tab:data}
\begin{tabular}{ c  c  c  c  c  c   c   c   c  c }
\hline
\\
Cluster & $\alpha_{2000}$  & $\delta_{2000}$  & $l$ & $b$
& $\mu_{\alpha}^{*}$ & $\mu_{\delta}$    &
$v_r$ & $d_{\odot}$ & [Fe/H]  \\
& ($^\circ$)& ($^\circ$)& ($^\circ$)& ($^\circ$)& (mas yr$^{-1}$) &(mas yr$^{-1}$) & (km s$^{-1}$)& (kpc) &
\\
\hline
\\
\\
Terzan 1 & 263.95 & -30.47 & 357.57 & 1.00 & $0.51\pm0.31$ &
$-0.93\pm0.29$ & $114.0\pm 14.0$ & $6.2\pm 0.6$ & -1.26\\
\\
Terzan 2  & 261.89 & -30.80 & 356.32 & 2.30 & $0.94\pm0.30$ &
$0.15\pm0.42$ & $109.0\pm15.0$ & $8.7\pm 0.8$ & -0.4\\
\\
Terzan 4 & 262.66 & -31.59 & 356.02 & 1.31 & $-3.50\pm0.69$ & $0.35\pm0.58$
& $-50.0\pm 2.9$ & $9.1\pm 0.9$ & -1.6 \\
\\
Terzan 9 & 270.41 & -26.84 & 3.61 & -1.99 & $0.00\pm0.38$ & $-3.07\pm0.49$
& $59.0\pm 10.0$ & $7.7\pm 0.7$ & -1.0 \\
\\
NGC 6522 & 270.89 & -30.03 & 1.02 & -3.93 & $3.35\pm0.60$ & $-1.19\pm0.34$
& $-21.1\pm 3.4$ & $7.8\pm 0.7$ &-0.95 \\
\\
NGC 6558 & 272.57 & -31.76 & 0.20 & -6.02 & $-0.10\pm0.55$ & $0.47\pm0.60$
& $-197.2\pm 1.5$ & $7.4\pm 0.7$ &-0.97 \\
\\
NGC 6540 & 271.53 & -27.76 & 3.29 & -3.31 & $0.07\pm0.40$ & $1.90\pm0.57$
& $-17.72\pm 1.4$ & $3.7\pm 0.3$ & -1.2\\
\\
Palomar 6 & 265.93 & -26.22 & 2.10 & 1.78 & $-3.27\pm0.41$ & $-1.44\pm0.19$
& $181.0\pm 2.8$ & $7.3\pm 0.7$ & -1.0 \\
\\
NGC 6652 \textsuperscript{b}& 278.94 & -32.99 & 1.53 & -11.38 & $-5.66\pm0.07$ & $-4.45\pm0.10$ & $-111.7\pm 5.8$ & $9.6\pm 0.9$ & -0.96\\
\\
\hline
\end{tabular}
 \textsuperscript{a}\footnotesize{Misprints in proper motions from Paper I have been corrected. These apply to Terzan 2, Terzan 4, NGC 6558, Palomar 6 and NGC 6652.}
 
 \textsuperscript{b}\footnotesize{Absolute proper motions taken from Sonh et al 2015.}
\end{table*}

\section{Methods}
\label{sec:methods}

We calculated the orbits of our nine  GCs by integrating them within both an axisymmetric and a barred mass model of the Galaxy. Our sample clusters are
 located in the inner region of the Galaxy, most of them inside $\sim 2$ kpc from the Galactic center. Therefore, a Galactic bar potential plays a key role in the dynamical analysis of these clusters.

\subsection{ A mass model of the Milky Way}

 In our analysis, we employ axisymmetric and nonaxisymmetric models for the Galactic gravitational potential.
According to observational evidence (Kent et al. 1991, Chemin et al. 2015), we modelled the axisymmetric bulge by adopting an exponential mass density profile. Such a profile is the S{\'e}rsic model. In particular, we adopted the representation given in Prugniel \& Simien (1997) (see also Terzi{\'c} \& Graham 2005), that closely matches the de--projected S{\'e}rsic $R^{1/n}$ profile
\begin{equation}
\rho_\mathrm{bulge} = \rho_0 \left( \dfrac{r}{R_\mathrm{e}}\right)^{-p}e^{-b (r/R_\mathrm{e})^{1/n}}
\end{equation}
where $\rho_0$ is the central density, $R_\mathrm{e}$ is the scale radius, $n$ is the curvature parameter, $b = 2n-1/3+ 0.009876/n$ for $0.5<n<10$ (Prugniel \& Simien 1997), and $p= 1.0 - 0.6097/n+0.05563/n^2$ for $0.6 < n< 10$. We refer to Terzi{\'c} \& Graham (2005) for more details. The exponential profile corresponds to the value of the curvature parameter $n=1$, which represents a pseudobulge with a density law similar to the one that is seen in the inner Galaxy (Freudenreich 1998). In the recent work of Chemin et al. (2015), the authors question whether the mass model of the Milky Way obtained by fitting the observed circular rotation curve in the inner 4--4.5 kpc of the Galaxy is correct or misinterpreted. One of their conclusions is that the
 velocity profile in the inner regions of the Galaxy does not represent the 
true rotation curve, but local motion. For our purposes, the most important result of
 their analysis is that the bulge of the Milky Way could be twice as massive and
 a half less concentrated than previously thought. Following their results, we assigned
 a total mass to the bulge equal to $M_\mathrm{bulge} = 1.2\times 10^{10} \;\;\; M_\odot$,
 and a scale radius equal to $R_\mathrm{e} = 0.87 \;\;\; \mathrm{kpc}$. This is in agreement
with the mass values given in Bland-Hawthorn \& Gerhard (2016) review.

In our Galactic mass model, the disc follows an exponential density profile with the form

\begin{equation}
\rho_\mathrm{disc} = \rho_0\exp(R/R_\mathrm{d})\exp(-|z|/h_\mathrm{z}).
\end{equation}

In general, the discs in galaxies are well represented by an exponential profile. Flynn et al. (1996) and more recently Smith et al. (2015) showed a method to approximate the gravitational potential generated by an exponential disc through the superposition of three Miyamoto--Nagai potentials
(Miyamoto \& Nagai 1975)
\begin{equation}
\Phi_\mathrm{disc} = \sum_{i=1}^3-\dfrac{G M_{\mathrm{d},i}}{\left\{x^2 + y^2 + \left[ a_{\mathrm{d},i} + (z^2 + b_{\mathrm{d},i}^2)^{1/2}\right]^2\right\}^{1/2}}
\end{equation}
where $M_{\mathrm{d},i}$ and $a_{\mathrm{d},i}$, $b_{\mathrm{d},i}$ are the masses and scale lengths of each component, respectively. 

The total disc mass of our model is $5.5 \times10^{10}$ M$_{\odot}$  with a radial exponential scalelength $R_\mathrm{d}=3.0$ kpc and exponential scaleheight of $h_\mathrm{z}= 0.25$ kpc (Bland-Hawthorn \& Gerhard 2016). The Miyamoto-Nagai parameters for the disc are listed in Table \ref{tab:parameters}. 

The dark matter halo is modelled with a NFW profile (Navarro et al. 1997). The parameters adopted  are shown in Table \ref{tab:parameters}.
\begin{equation} 
\rho_\mathrm{h}(R)=\dfrac{M_\mathrm{h}}{4\pi} \dfrac{1}{(a_\mathrm{h} + R)^2R} \;\;\;,
\end{equation}
where $M_\mathrm{h}$ is the total mass of the halo component and $a_\mathrm{h}$ is a scalelength.

\begin{figure}
\hskip -5mm
\includegraphics[scale=0.5]{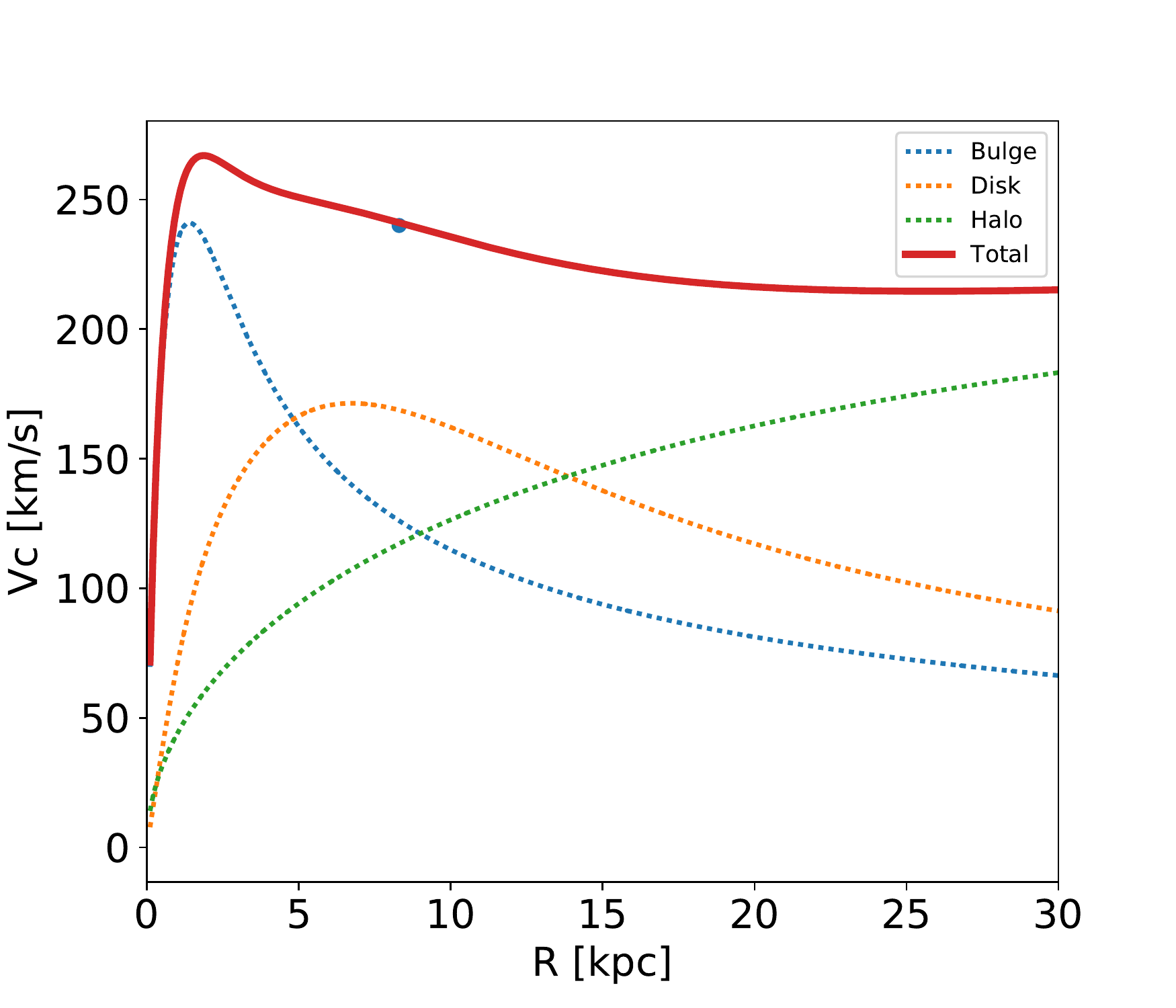}
\caption{Total circular velocity curve and the contribution of each component associated to the axisymmetric Galactic mass model. The blue dot shows the velocity at the Sun position.}
\label{fig:velocity_curve}
\end{figure}

Figure \ref{fig:velocity_curve} shows the circular velocity curve associated with the adopted axisymmetric Galactic mass model and the contribution of each component to the circular velocity. In our model the Sun is placed at 8.33 kpc with a circular velocity of 241 km s$^{-1}$.

 The presence of a bar in the inner regions of the Milky Way is well-known 
(e.g. Binney et al. 1991; Weiner \& Sellwood 1999, Hammersley et al. 2000,
  Benjamin et al. 2005) - see a more detailed discussion in Barbuy et al. (2018, and references therein). 
In order to go beyond a simplistic axisymmetric modelling of the central regions of the Galaxy, we added a single rotating bar to our mass model. 
Several models of the Galactic bar have been proposed in the past (e.g. Long \& Murali 1993, Pichardo et al. 2004). In the present work we 
adopted the widely used (e.g. Pfenniger 1984, Weiner \& Sellwood 1999, Gardner \& Flynn 2010, Berentzen \& Athanassoula 2012,
 J{\'{\i}}lkov{\'a} et al. 2012, Casetti-Dinescu et al. 2013) description in terms of a triaxial rotating Ferrer's ellipsoid 
\begin{equation}
\rho_\mathrm{bar}=
\begin{cases}
\rho_{\mathrm{c}}(1-m^2)^n\;\;\;\; & ,\;m< 1\\
0 & ,\; m\geq 1
\end{cases}
\label{eq:ferrers}
\end{equation}
where $\rho_\mathrm{c}$ is the central density, $n$ is a positive integer and 
\begin{equation}
m^2=\dfrac{x^2}{a^2}+\dfrac{y^2}{b^2}+\dfrac{z^2}{c^2}\;\;\; ,
\end{equation}
with $a$, $b$, ${c}$ the semiaxes of the ellipsoid. The adopted value of the parameters
 of the bar and the reference papers are summarized in Table \ref{tab:parameters}.
 When adding the bar to the mass model,  all the mass in the S\'ersic component is employed to build the Galactic bar.

In our calculations we neglected the presence of spiral arms perturbing the gravitational potential of the Galactic disc. The reason
 for this is that the clusters in our sample are confined within the inner $\simeq 4$ kpc of the Galaxy, which is approximately the size of the semi-major axis of the Galactic bar. We then do not expect the orbits to be perturbed by a rotating 
spiral pattern, which typically develops outside the volume occupied by the bar.

\begin{table}
\caption{Parameters of the adopted Galactic mass model.}
\label{tab:parameters}
\begin{tabular}{l c c}
\hline
\\
Parameter&Value & Reference\\
\hline
\multicolumn{3}{c}{\textbf{Axisymmetric components}}\\
\hline
\multicolumn{3}{c}{\textbf{Bulge}}\\  
 $M_\mathrm{b,tot}$ &$1.2 \times 10^{10}$ $M_\odot $ & 1, 2 \\
 $R_\mathrm{e}$ & $0.87$ $\mathrm{kpc}$ &  \\
 $n$ &$1$ & 3\\
\\
\multicolumn{3}{c}{\textbf{Disc}}\\
$M_\mathrm{d,1}$ & $1.07 \times 10^{11}$ $M_\odot $ & \\
$a_\mathrm{d,1} $ &$5.97$ $\mathrm{kpc}$ &\\
$M_\mathrm{d,2}$ & $-7.09 \times 10^{10}$ $M_\odot $ & \\
$a_\mathrm{d,2} $ &$12.99$ $\mathrm{kpc}$ &\\
$M_\mathrm{d,3}$& $1.2 \times 10^{10}$ $M_\odot $ & \\
$a_\mathrm{d,3} $ & $2.04 \;\mathrm{kpc}$ &\\
$b_\mathrm{d}$ & $0.25\; \mathrm{kpc}$ &\\               
\\
\multicolumn{3}{c}{\textbf{Halo}}\\
$M_\mathrm{h} $ & $3.0 \times 10^{12} \; M_\odot $ &  4\\ 
$a_\mathrm{h}$ & $57.0 \; \mathrm{kpc}$ & 4\\
\\
\hline
\multicolumn{3}{c}{\textbf{Galactic bar}}\\
\hline
 $M_\mathrm{bar}$ & $1.2 \times 10^{10}$ $M_\odot $ & 1, 2, 5\\
 $n$ & $2$ & 6\\
 $a$ & $3.5 \; \mathrm{kpc}$ &9, 10 \\
 $b$ & $1.4 \; \mathrm{kpc}$ & 9, 10\\
 $c $ &$1.0 \; \mathrm{kpc}$ &9, 10 \\
 $\phi_\mathrm{bar}$ & $25^\circ$  & 2,7, 8 \\ 
 $\Omega_\mathrm{bar} $ &$40, 50, 60$ $\mathrm{km} \,\mathrm{s}^{-1}\, \mathrm{kpc}$ &2, 11, 12 \\
\hline
\end{tabular}
\\
{\footnotesize{ \textbf{References.} (1) Portail et al. 2015; (2) Bland-Hawthorn \& Gerhard 2016; (3) Kent et al. 1991; (4) Irrgang et al. 2013; (5) Weiner \& Sellwood 1999; (6) Pfenniger 1984; (7) Rattenbury et al. 2007; (8) Wegg \& Gerhard 2013; (9) Freudenreich 1998; (10) Gardner \& Flynn (2010); (11) Bissantz et al. 2003; (12) Portail et al. 2017.}}
\end{table}

\subsection{Integration of the orbits} \label{sec:integration}
We integrated the Galactic orbits of the globular clusters using \textsc{nigo} (Numerical Integrator of Galactic Orbits --Rossi 2015a), which is ideally suitable for the purposes of our analysis. The code includes an implementation of the mass model that we adopted in this work. The solution of the equations of motion is evaluated numerically using the Shampine--Gordon integration scheme (for further details we refer to Rossi (2015b)). We integrated the orbits of our sample of star clusters forward in time for 10 Gyr. Using the observational parameters of the GCs given in Table \ref{tab:data}, we computed the initial state vector of the clusters assuming the Sun's Galactocentric distance $R_0=8.33$ kpc and a circular velocity $V_0=241$ km s$^{-1}$ (e.g. Bland-Hawthorn \& Gerhard 2016, and references therein). The velocity components of the Sun with respect to the local standard of rest are $(U,V,W)_{\odot}= (11.1, 12.24, 7.25)$ km s$^{-1}$ (Shc{\"o}nrich et al. 2010). We defined as inertial Galactocentric frame of reference the right--handed system of coordinates ($x,y,z$) where the $x$--axis points to the Sun from the Galactic Centre and the $z$--axis points to the North Galactic Pole. We refer to the bar--corotating frame of reference as the right--handed system of coordinates ($x_\mathrm{r},y_\mathrm{r},z_\mathrm{r}$) that co--rotates with the bar, where the $x_\mathrm{r}$ axis is aligned with the bar semi--major axis $a$ and $z_\mathrm{r}$ points towards the North Galactic Pole. As we mention in Section \ref{sec:data}, our proper motions are relative to the bulge field stars, therefore we have to give a special treatment in order to get the velocity vector, as was explained in Terndrup et al. (1998) and in Paper I. 
It is worth to point out that Bobylev \& Bajkova (2017) carried out an orbital study with the same sample GCs and data from Paper I.
However, they did not apply the correct conversion from relative proper motions into velocities, assuming that the proper motions
given in Paper I were absolute. Therefore their calculated velocity vectors are not consistent with this sample of bulge GCs. 
For NGC 6652, the proper motions are absolute, such that for this cluster we apply the usual way to convert the proper motions into velocities.

\section{Orbital properties of the bulge globular clusters}
\label{sec:cluster_orbits}

In this Section we present the main properties of the Galactic orbits determined for each of the nine globular clusters in our sample, both for the cases of an axisymmetric and a barred Galactic mass model using three different values of the angular velocity of the bar $\Omega_b= 40$, 50, and 60 km s$^{-1}$ kpc$^{-1}$. 

\begin{figure*}
\begin{tabular}{l l}
\includegraphics[scale=0.17]{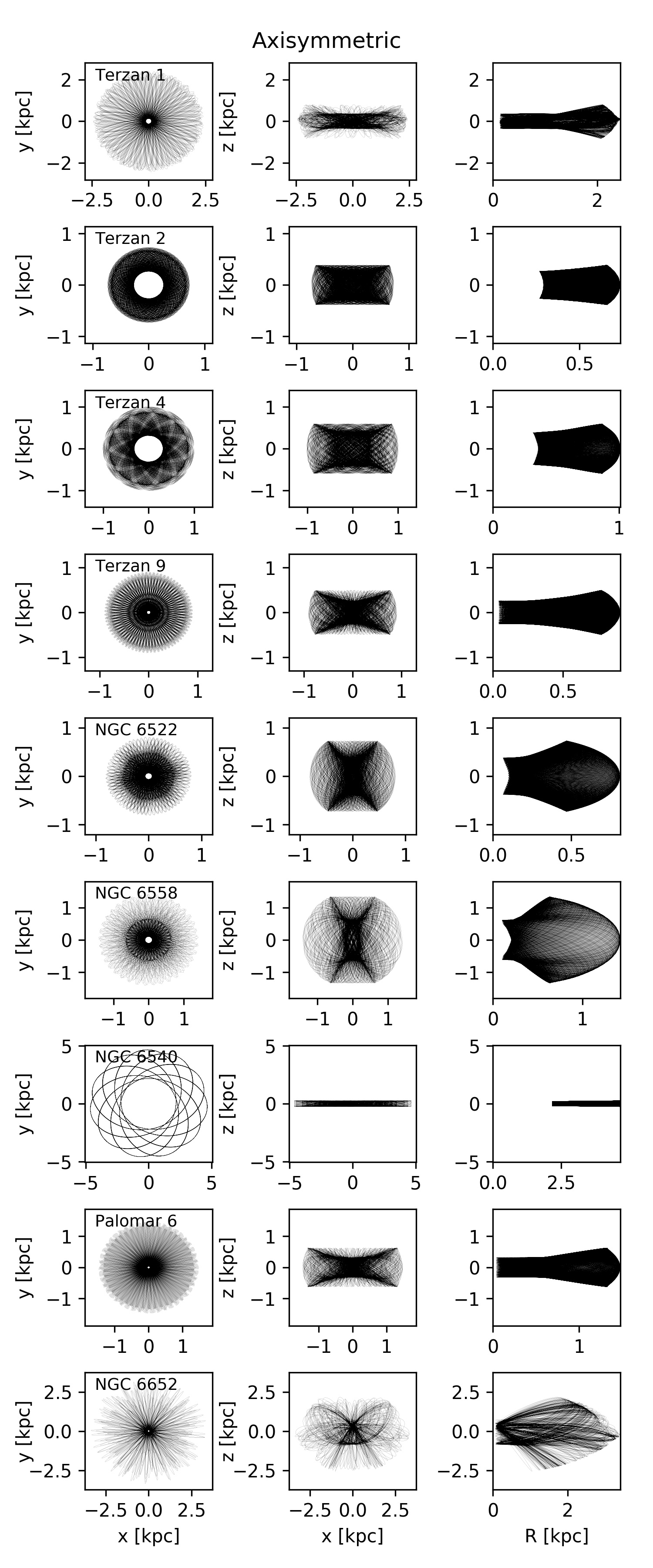} &
\includegraphics[scale=0.17]{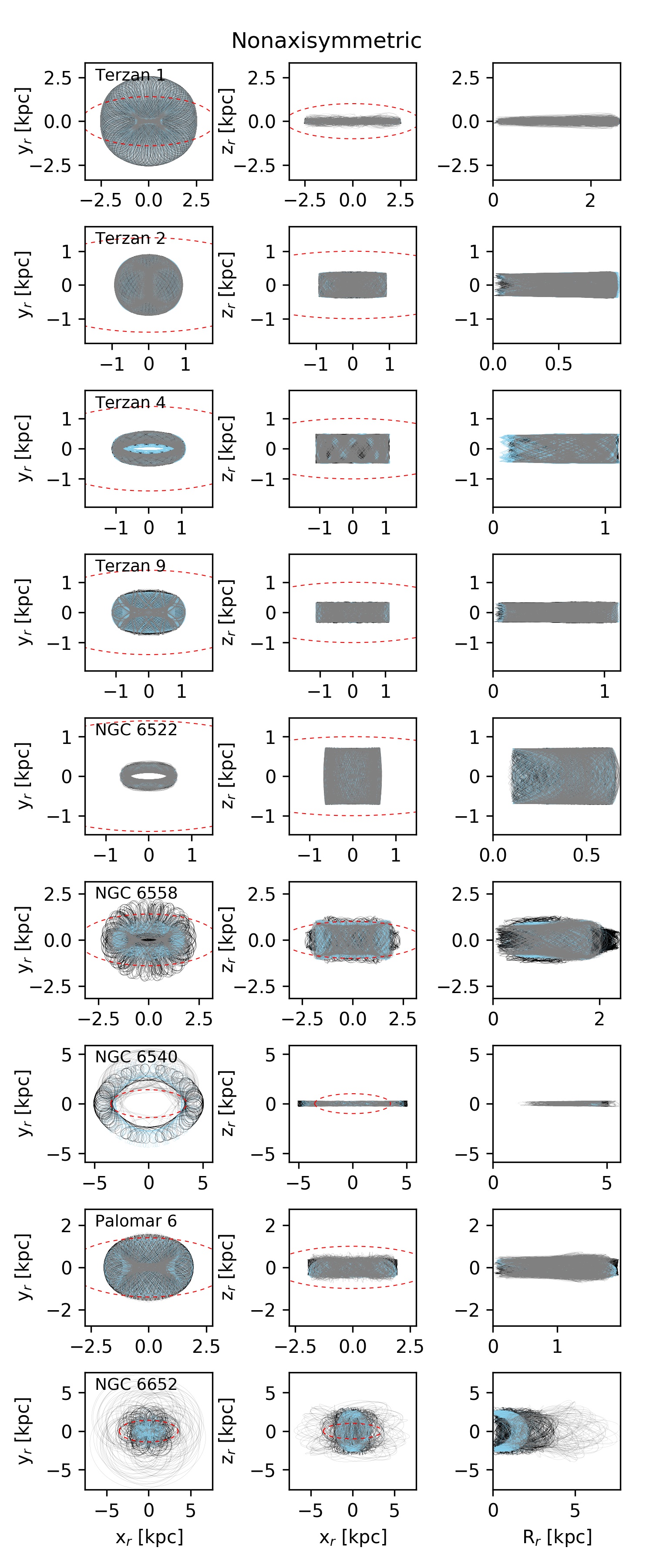} 
\end{tabular}
\caption{Orbits for the sample of globular clusters. The three left columns show $x-y$, $x-z$, and $R-z$ projections for orbits with the axisymmetric Galactic potential, while the three right columns show the orbits with the nonaxisymmetric Galactic potential co-rotating with the bar. The colors in the left panels are the orbits with different pattern speed of the bar, 40 (blue), 50 (black), and 60 (grey) km s$^{-1}$ kpc$^{-1}$. The dashed red line shows the size of the Galactic bar.}
\label{fig:orbits}
\end{figure*}

 Figure \ref{fig:orbits} shows the orbit for each individual globular cluster using as initial condition the
 central values presented in Table \ref{tab:data}, both in the case of the axisymmetric potential (three left panels) and the model with bar (three right panels). For each cluster, we plotted the projection of the orbit on the Galactic plane ($x-y$), on the ($x-z$) plane and on the meridional plane ($R-z$). The orbits obtained assuming an axisymmetric potential are shown in the inertial Galactic frame of reference, while the orbits obtained in the barred potential are shown in the bar co--rotating frame of reference. In the model with bar, we use three different
 angular velocities, that are plotted with different colors in the right panels of Figure \ref{fig:orbits}.

For each orbit in Figure \ref{fig:orbits}, we calculate the perigalactic distance $r_{\mathrm{min}}$, the apogalactic distance $r_{\mathrm{max}}$, the maximum vertical excursion from the Galactic plane $|z|_{\mathrm{max}}$, and the eccentricity  defined as $e=(r_{\mathrm{max}}- r_{\mathrm{min}} )/(r_{\mathrm{max}}+ r_{\mathrm{min}} )$. In order to know whether the orbital motion of globular clusters has a prograde or a retrograde sense with respect to the rotation of the bar, we calculate the z-component of the angular momentum in the inertial frame, $Lz$. Since this quantity is not conserved in a model with nonaxisymmetric structures, we are interested only in the sign. 

We can see that with the axisymmetric potential, most of clusters are confined inside $\sim 2$ kpc in the Galactic plane, except for NGC 6540 and NGC 6652, that reach distances between $\sim 4$ and 5 kpc. The clusters have a z--distance from the Galactic plane of $\sim 0.6 $, except for NGC 6558 with $z\sim 1.3$, and NGC 6652 with $z\sim 2.6$ kpc. Most of clusters show high eccentricity $e> 0.8$, although there are three clusters with $e<0.5$ (Terzan 2, Terzan 4, and NGC 6540). 

When the bar is introduced to the model, there is an interesting dynamical effect produced by this structure. In general, the clusters are confined in the bar region.
 For the cases of Terzan 1 and NGC 6540, these clusters are going inside and outside of the bar in the Galactic plane, with low vertical excursions from the plane.
 It is not surprising that NGC 6652 is also completely outside of the bar region because we know that this cluster belongs to the halo component. 
We can also notice that Terzan 4, Terzan 9, NGC 6522, NGC 6558, Palomar 6 have a bar-shape orbit in the ($x_r-y_r$) projection, and a boxy-shape in the ($x_r-z_r$),
and this means that these clusters are trapped by the bar.  NGC 6540 has a particular and different behaviour as compared with the other sample clusters:
 it is confined to the Galactic plane, and has the correct energy to go inside and outside of the bar,
 and the orbit is trapped by a higher-order resonance or corotation, depending on the bar angular velocity, $\Omega_b$. 
With respect to the different values of $\Omega_b$, we can see that most of orbits are not sensitive to the change of this parameter, except for NGC 6558, NGC 6540, and NGC 6652. Regarding the sense of the orbital motion, we have clusters with prograde orbits
 such as Terzan 4, NGC 6522, NGC 6558, NGC 6540, and we have clusters that have prograde and retrograde orbits at the same time. The prograde-retrograde clusters are Terzan 2, Terzan 9, Palomar 6 and NGC 6652. 
The prograde-retrograde orbits have been observed before in others globular clusters (Pichardo et al. 2004; Allen et al. 2006). This effect is produced by the presence of the bar, and it could be related to chaotic behavior, however this is not well-understood yet.

\subsection{Orbital properties from the Monte Carlo method}

In order to evaluate the uncertainties associated with the parameters of
 the clusters, we followed a Monte Carlo method. For each cluster we generated a set of 1000 initial conditions included within the errors affecting the values of heliocentric distance, proper motions and heliocentric radial velocity, presented in Table \ref{tab:data}. We do this with the purpose of seeing how much the orbital properties of the clusters such as the perigalactic distance, the apogalactic distance, the maximum vertical height and the eccentricity change, considering the errors of the measurements, and at the same time, how the Galactic bar affects them.
We integrated this set  of orbits for 10 Gry using \textsc{nigo} and evaluated the mean value and standard deviation of the orbital parameters.
 
In Table \ref{tab:orbital_parameters}, we give for each globular cluster the orbital parameters obtained with the axisymmetric potential (first line) and the model with bar, where we employed three values of $\Omega_b= 40$ (second line), $50$ (third line) and $60$ (forth line) km s$^{-1}$ kpc$^{-1}$. Columns 2, 3, and 4 show the average perigalactic distance, the average apogalactic distance, and the average maximum vertical excursion from the Galactic plane, respectively. Column 5 gives the average orbital eccentricity. The errors provided in each column are the standard deviation of the distribution. 

\begin{figure*}
\includegraphics[scale=0.20]{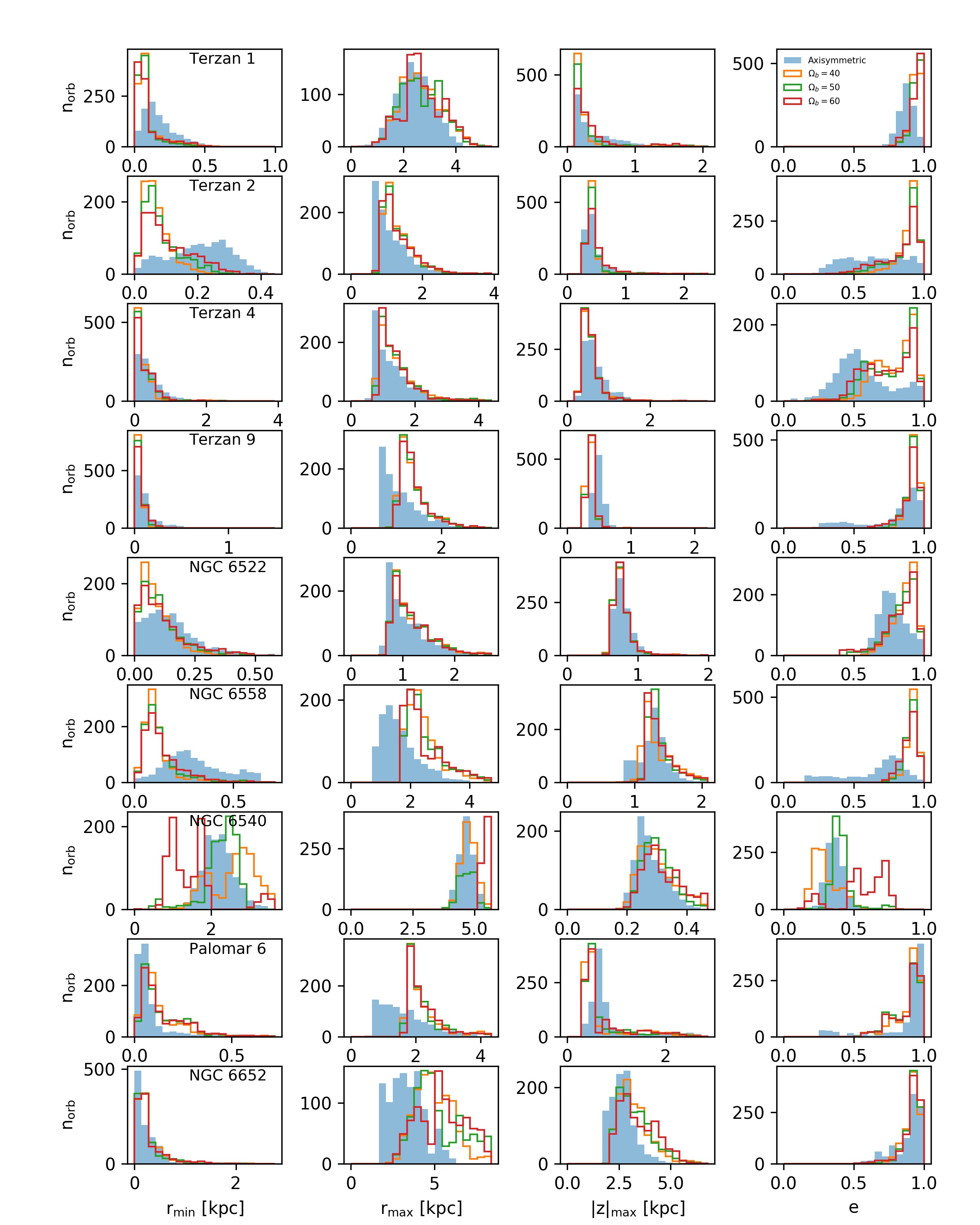} 
\caption{Distribution of orbital parameters for each cluster. The distribution for perigalactic $r_{\mathrm {min}}$, apogalactic $r_{\mathrm {max}}$, maximum vertical hight of orbit $|z|_{\mathrm{max}}$, and eccentricity $e$. Different colors show the distribution for the axisymmetric model (blue), model with bar using angular velocity $\Omega_b=$ 40 (orange), 50 (green), and 60 (red) km s$^{-1}$ kpc$^{-1}$.}
\label{fig:hist}
\end{figure*}

Figure \ref{fig:hist} shows the distribution of each orbital parameter, for the axisymmetric model (blue)  and the model with bar using a $\Omega_b= 40$ (orange), 50 (green), and 60 (red) km s$^{-1}$ kpc$^{-1}$. The clusters have perigalactic distances $r_{\mathrm{min}}<1$ kpc with the axisymmetric model, and these decrease with the presence of the bar, $r\mathrm{min}<0.5$ kpc.
 There is a particular cluster, NGC 6540, with perigalactic distance with the axisymmetric model between 1 and 3 kpc, and with bar model between 0.5 and 4 kpc, 
depending on the pattern speed. Regarding the apogalactic distance, the clusters can reach maximum distances up to 4 kpc, and there are two clusters that could go further, NGC 6540 (up to 6 kpc) and NGC 6652 (up to 8 kpc). As for the maximum height, the clusters could go between 0.5 to 1.5 kpc, and for NGC 6652 up to 5 kpc. With respect to eccentricity, most clusters have high eccentricities, and the bar model makes the distribution in eccentricities to be narrower than with the
axisymmetric model. The exceptions are Terzan 4, 
with a low eccentricity with no bar, whereas the orbits become highly eccentric with the presence of the bar; and
NGC 6540,  with a very low eccentricity, $e<0.5$, with the axisymmetric model, whereas with the bar model
 the eccentricity depends on the angular velocity, higher if $\Omega_b>50$ km s$^{-1}$ kpc$^{-1}$ or lower if $\Omega_b<50$ km s$^{-1}$ kpc$^{-1}$.

From the average values given in Table \ref{tab:orbital_parameters}, we can see that the Galactic bar induces to have smaller average perigalactic distances (except for NGC 6540, Palomar 6 and NGC 6652), larger average apogalactic distances, lower vertical excursions from the Galactic plane (except for NGC 6652), and higher eccentricities (except for NGC 6540). The effect produced by different pattern speeds on the clusters are almost negligible, except for NGC 6540.

\begin{table*}
\caption{Orbital parameters with the Axisymmetric and Nonaxisymmetric Potentials.}
\label{tab:orbital_parameters}
\centering
\scalefont{0.85}
\begin{tabular}{c c c c c  c}
\hline
\\
Cluster & $<r_\mathrm{min}>$  & $<r_\mathrm{max}>$  & $<|z|_\mathrm{max}>$  & $<e>$ & [Fe/H]\\
             & (kpc) &  (kpc)& (kpc) &  \\
\\
\hline
\\

Terzan 1 & $0.185 \pm 0.125$ &$2.453 \pm0.685 $&$0.446 \pm 0.363$ &$0.870 \pm 0.065$ & -1.26\\
	     &$0.095 \pm 0.092$& $2.672 \pm 0.773$& $0.280 \pm 0.322$& $0.937 \pm 0.038$ &\\ 
	     &$0.086 \pm 0.084$ & $2.690 \pm 0.793 $&$ 0.290 \pm 0.304$ & $0.943 \pm 0.034$ &\\
	     &$0.097 \pm 0.109 $&$ 2.731 \pm 0.774$ & $0.371 \pm 0.363$ &$0.940 \pm 0.043$ &\\
\\        
Terzan 2 & $0.206\pm 0.095 $& $1.100\pm 0.461$ & $0.467\pm 0.215$ &$ 0.642 \pm0.204$&  -0.4\\
	      & $0.070 \pm 0.040$& $1.318 \pm 0.439$& $0.433 \pm 0.169 $&$ 0.892 \pm 0.072 $ & \\        
	      &$0.084 \pm 0.056$ & $1.313 \pm 0.448 $&$ 0.445 \pm 0.188 $& $0.867 \pm 0.108 $& \\
	     &$0.106 \pm 0.072 $&$1.331 \pm 0.521 $& $0.489 \pm 0.223$ & $0.834 \pm 0.135$&\\
\\
Terzan 4 &$0.442\pm 0.431 $&$1.272 \pm0.567 $&$0.689\pm 0.308$ & $0.562\pm 0.193$& -1.6\\
	      &$0.239 \pm 0.350$ & $1.482 \pm 0.695$ & $0.585 \pm 0.300 $& $0.787 \pm 0.145 $&\\
	      &$0.257 \pm 0.336 $& $1.468 \pm 0.683$ & $0.579 \pm 0.297$ & $0.767 \pm 0.158 $&\\
	      &$0.276 \pm 0.320 $& $1.477 \pm 0.755$ & $0.581 \pm 0.297$ & $0.741 \pm 0.165 $&\\         
\\
Terzan 9 &$0.129 \pm0.146$ &$1.108 \pm0.439$ &$0.506\pm 0.146$ &$0.784 \pm0.208$ & -1.0\\
          	&$0.056 \pm 0.039 $& $1.388 \pm 0.348 $&$ 0.365 \pm 0.088$ & $0.923 \pm 0.039 $&\\
	      &$0.062 \pm 0.049$ & $1.394 \pm 0.349 $& $0.370 \pm 0.085$ &$ 0.916 \pm 0.046 $&\\
	      &$0.073 \pm 0.065$ & $1.412 \pm 0.364 $& $0.375 \pm 0.088$ & $0.905 \pm 0.067 $&\\         
\\
NGC 6522 & $0.149 \pm 0.102$ &$1.066 \pm0.365$& $0.801 \pm 0.152 $&$0.776 \pm 0.096 $& -0.95\\
        		&$0.089 \pm 0.062$ &$ 1.176 \pm 0.386$ &$ 0.772 \pm 0.140$ &$ 0.863 \pm 0.080 $&\\
	      &$0.109 \pm 0.085$ & $1.178 \pm 0.391$ & $0.772 \pm 0.136$ & $0.838 \pm  0.103 $&\\
	      &$0.117 \pm 0.097$ & $1.187 \pm 0.397$ & $0.777 \pm 0.138$ &$ 0.829 \pm 0.121$ &\\         
\\
NGC 6558 & $0.308 \pm 0.148$ &$1.650\pm 0.632$ &$1.273\pm 0.209$& $0.654 \pm0.213$& -0.97\\
  		&$0.114 \pm 0.078$ & $2.466 \pm 0.608 $& $1.375 \pm 0.273$ & $0.914 \pm 0.043$ &\\
	      &$0.129 \pm 0.095$ & $2.527 \pm 0.685 $& $1.384 \pm 0.221$ & $0.908 \pm 0.045$ &\\
	      &$0.141 \pm 0.099$ &$ 2.563 \pm 0.892 $& $1.381 \pm 0.227$ & $0.897 \pm 0.054$ &\\         
\\
NGC 6540 & $2.205 \pm0.394$ &$4.674 \pm0.305$&$ 0.279 \pm0.049$& $0.364 \pm0.056$ & -1.2\\
               &$2.627 \pm 0.606$ & $4.773 \pm 0.319$ & $0.313 \pm  0.100$ &$ 0.299 \pm 0.091$ &\\
	      &$2.264 \pm 0.490 $& $5.348 \pm 0.726 $& $0.324 \pm 0.115$ &$ 0.411 \pm 0.083 $&\\
	      &$1.628 \pm 0.866 $&$ 5.615 \pm 0.229 $&$ 0.348 \pm 0.111 $&$ 0.567 \pm 0.168 $&\\         
\\
Palomar 6 & $0.089\pm 0.113$& $1.624 \pm0.738$ &$0.836 \pm0.516$ &$0.866 \pm0.190$ & -1.0\\
  		 &$0.129 \pm 0.111$ &$ 2.237 \pm 0.628 $& $0.701 \pm 0.541$ &$ 0.888 \pm 0.092$ &\\
	      &$0.147 \pm 0.118$ & $2.269 \pm 0.665 $& $0.692 \pm 0.493 $& $0.871 \pm 0.102 $ &\\
	      &$0.147 \pm 0.121$ & $2.353 \pm 0.795 $& $0.737 \pm 0.527$ & $0.876 \pm  0.101$ &\\         
\\
NGC 6652 & $0.220 \pm0.218$ & $3.381 \pm1.035$ &$2.693 \pm0.589$& $0.882 \pm0.102 $& -0.96\\
  		 &$0.258 \pm 0.252$ &$ 5.056 \pm 1.250$ &$ 3.221\pm  0.762$ &$ 0.910 \pm 0.057$ &\\
	      &$0.263\pm  0.288$ & $5.358 \pm 1.616$ & $3.288 \pm 0.815$ &$ 0.916 \pm 0.059$&\\
	      &$0.291 \pm 0.316$ & $5.769 \pm 1.859$ &$ 3.459 \pm 0.911$ &$ 0.913 \pm 0.068$ &\\         
\\     
\hline 
\end{tabular}
\end{table*}

\section{Correlations between orbital parameters and chemical properties}
\label{sec:correlations}

After determining the Galactic orbits of the clusters in our sample, we combined the information on the  orbital 
properties with the metallicity, [Fe/H], given that we are particularly interested in evaluating any obvious dependence 
among them. In fact, we already know from Paper I that the spatial distribution of the globular clusters in the Galaxy has a strong dependence on their metallicity. We now aim to refine this analysis for the bulge clusters of the present sample.

\begin{figure}
\centering
\includegraphics[scale=0.8]{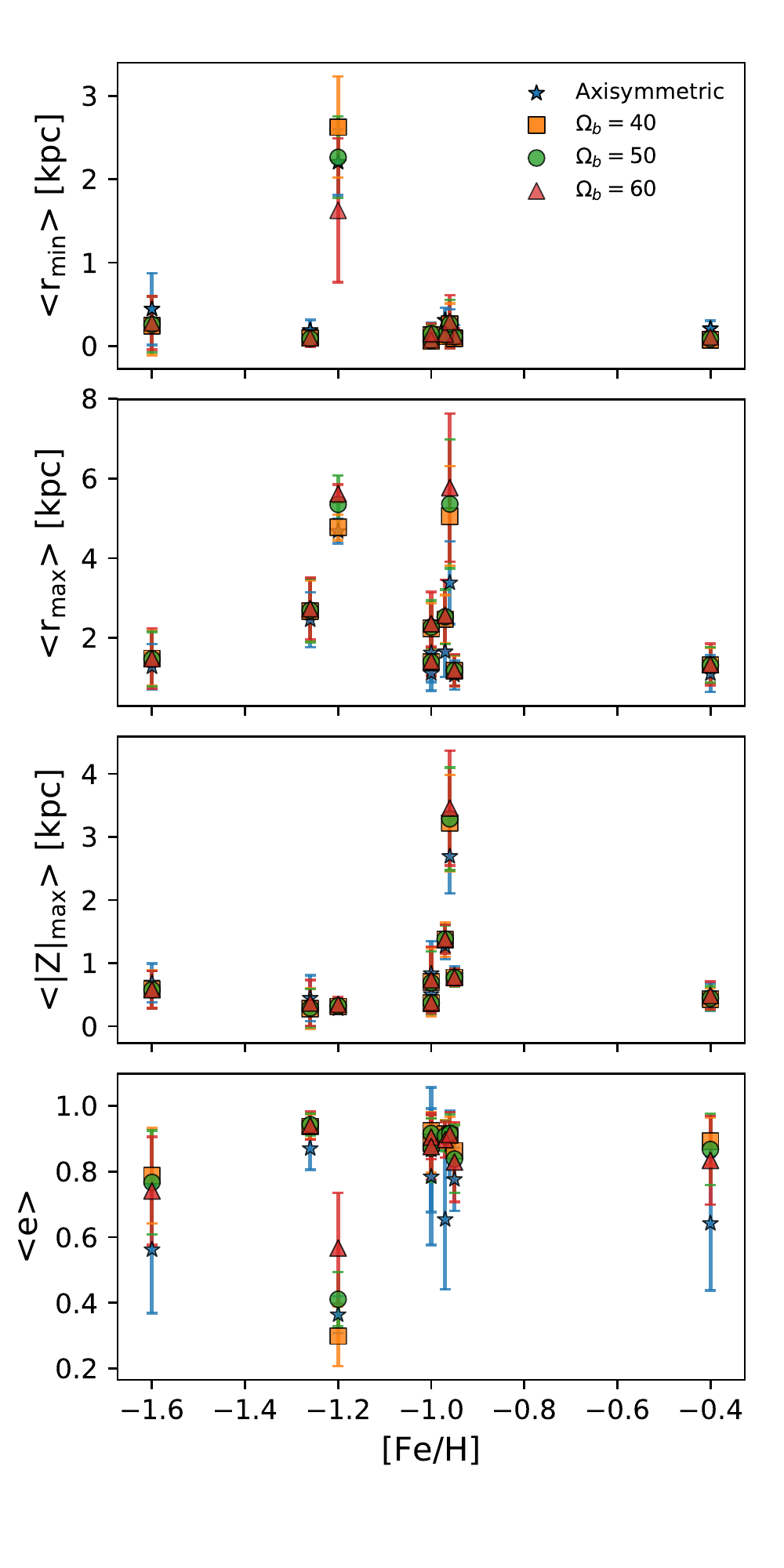}
\caption{Scatter plots of metallicity versus orbital parameters. Read from the top. First panel: the average perigalactic distances. 
Second panel: the average apoigalactic distances. Third panel: the average maximum excursion to the Galactic plane. Fourth panel: the average eccentricities. The blue stars show the results obtained for the axisymmetric mass model, while the orange squares, green circles and red triangles show the results obtained for the model with bar using $\Omega_b=$ 40, 50, and 60 km s$^{-1}$ kpc$^{-1}$, respectively.}
\label{fig:Fe_H_corr}
\end{figure}

 We constructed scatter plots of the orbital parameters as function of the clusters' metallicities. Figure \ref{fig:Fe_H_corr} shows the distribution of perigalactic distance, apogalactic distance, maximum height from the Galactic plane, and orbital eccentricity of the clusters as function of their metallicity. We studied both cases of an axisymmetric model (blue stars) and a barred mass model with $\Omega_b=40$ (orange squares), 50 (green circles), and 60 (red triangles) km s$^{-1}$ kpc$^{-1}$. We can see that the clusters with [Fe/H]$\gtrsim -1.0$ have similar values in all orbital parameters. If we consider only the clusters with [Fe/H]$<-1.0$, the perigalactic distance and the height from the Galactic plane
might tend to decrease with metallicity, while the apogalactic distance and eccentricity increase.
NGC 6540  has a completely different behaviour in comparison with the other clusters. Also we expect NGC 6652
([Fe/H]$\sim-1.0$) having different orbital parameters from the rest, because it belongs to the halo component. 

\begin{figure*}
\centering
\includegraphics[scale=0.70]{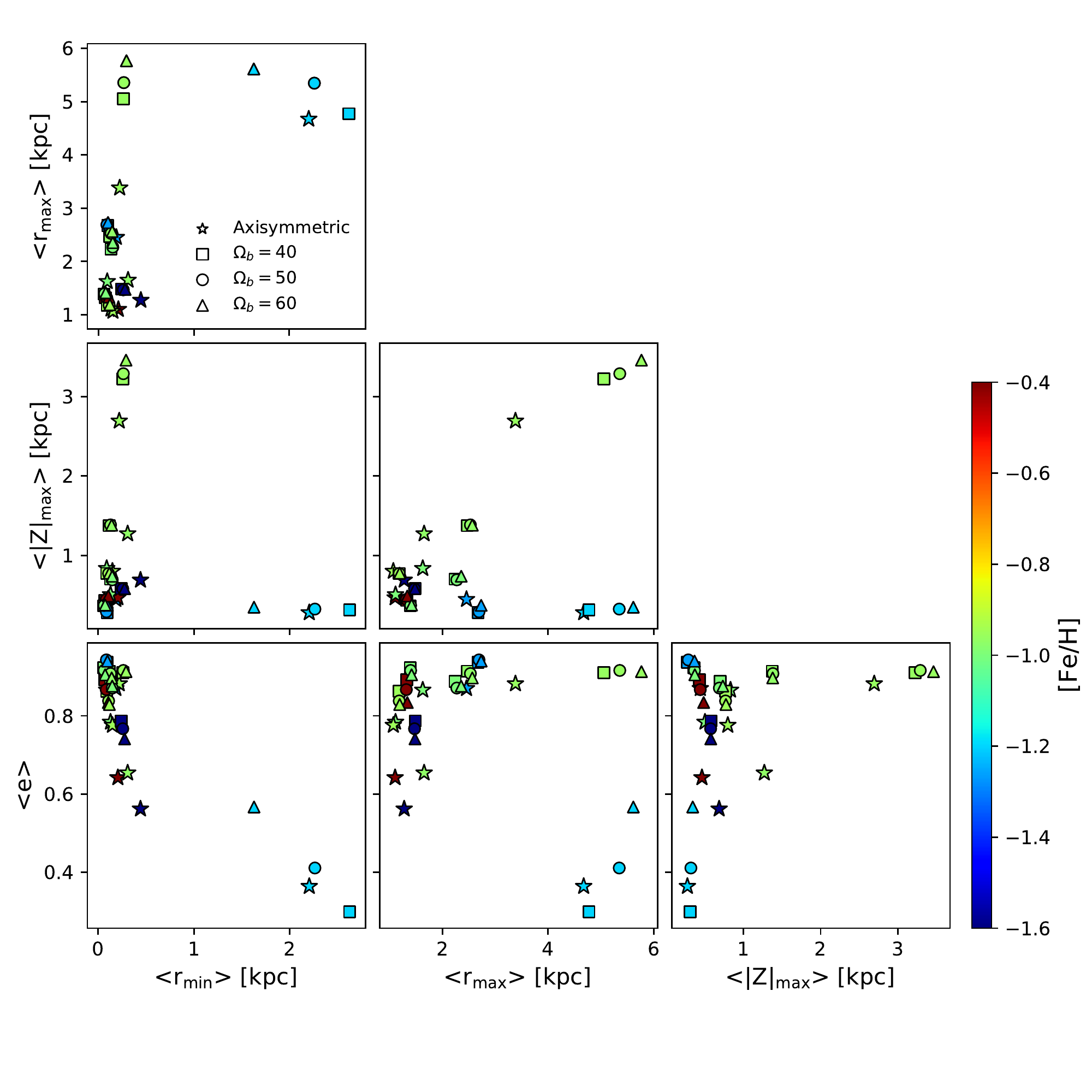}
\caption{Orbital parameters as function of average perigalactic distance $<r_{min}>$, average apogalctic ditance $<r_{max}>$, average of maximum distance from the Galactic plane $<|z|_{max}>$, and average eccentricity $e$, for the axisymmetric model (stars), model with bar and $\Omega_b=$ 40 (squares), 50 (circles), and 60 (triangles) km s$^{-1}$ kpc$^{-1}$. The color bar is the cluster metallicity [Fe/H].}
\label{fig:Dy_e_fe}
\end{figure*}

 In addition, we looked for any correlation among the different orbital parameters of the clusters. We expect the clusters sharing their dynamical
 properties with some dispersion, due to being confined in the bulge/bar region. 
In Figure \ref{fig:Dy_e_fe}, we show a scatter plot of all the possible combinations among the orbital parameters, where the colour represents the metallicity. 
We see that, in each panel, most of clusters are gathered in the same region indicating that they belong to the same component, 
in this case that they are confined to the bulge/bar. On the other hand,  NGC 6652 and NGC 6540 have a different behaviour relative to the other ones.
 We know that NGC 6652 belongs to the halo component, therefore it does not have to share orbital properties with the bulge/bar globular clusters,
 and NGC 6540 has properties consistent with the thick disc kinematics.

\section{Summary and Conclusions}
\label{sec:discussion}

 We have employed the available data of the bulge globular clusters given in Paper I to compute their orbits, and compare the orbital properties with their metallicity.
 This analysis has been carried out both with an axisymmetric potential and a model with the Galactic bar, where we vary the angular velocity of the bar. In order to take into account the effect of the observational uncertainties, we employed the Monte Carlo method to construct a set of 1000 initial conditions for each cluster.

The clusters in our sample, which are projected onto the Galactic bulge, show trajectories indicating that they are confined to a bulge/bar. The only exception applies to
the comparison cluster NGC 6652, which shows an orbit expected from an inner halo cluster. We found that most of clusters in the inner 4 kpc from the Galactic centre exhibit high eccentricities with close perigalactic passages. However we also  identified an object with orbital properties completely different compared with the bulge/bar and halo component, characterised by a disc-like orbit (NGC 6540), having a larger perigalactic distance larger than 5 kpc with very low eccentricities, $e<0.4$. The maximum heights with respect to the Galactic plane for all the sample clusters (except for NGC 6652), are typically below 1.5 kpc. It appears that the sample clusters evolve within the region occupied by the Galactic bar. More specifically, due to the gravitational effect produced by the
 bar, the clusters could be trapped by some resonance of the bar, and
in this case they would
 be those that follow a bar-shape in $(x_r-y_r)$ and $(x_r-z_r)$, in the rotating frame. On the other hand, the cluster Terzan 1, that even though it is confined in the inner region,
 its orbit does not have a bar/boxy-shape, therefore this cluster is probably not trapped by the bar. In any case,
 as expected, the presence of a rotating bar in the centre of the Galaxy has a strong impact on the orbital evolution of the inner clusters, although the effect produced by different angular velocities on the clusters seems to be almost negligible, except for NGC 6540.

We found that most clusters in the inner $\sim$ 3 kpc  are part of the bulge/bar population. By this we mean that either these clusters were formed early on before bar formation, and were trapped
during the bar formation, or else, they formed together with the bar.  On the other hand, it appears that in the region outside
 the bar, the clusters appear to behave as thick disk-like orbits or halo, such as NGC 6540 and NGC 6652, respectively.
 
 With respect to metallicity correlations,  the properties are similar for clusters with [Fe/H]$\gtrsim -1.0$. 
On the other hand, if we consider only clusters with [Fe/H]$<-1.0$, the perigalactic distance and the maximum high excursion from the Galactic plane
tend to decrease with metallicity, while the apogalactic and eccentricity increase. However, there is not any obvious dependence among the metallicity and orbital properties. We suggest that a more extensive analysis of  a larger bulge cluster sample is required to improve the statistical significance of these results.

 Our results could provide some insight on the formation processes of the globular clusters 
located in the Galactic bulge. Considering the old nature of these objects (e.g. 
Ortolani et al. 2011; Kerber et al. 2018), they were formed during the very early stages of the evolution of the Milky Way. 
Also, many of the sample clusters belong to the metal--rich Galactic sub--population (Paper I and references
 therein) and therefore they likely formed within high--pressure, metal--rich gas Galactic structures at 
high redshift. 

 In our analysis, we have orbits with prograde motions and orbits that change their sense of motion from prograde to retrograde,
 as seen from the inertial frame. The prograde-retrograde orbits could be related to chaotic behavior (Pichardo et al. 2004; Allen et al. 2006). The absence of strong correlations between orbital
 and chemical properties of the clusters could be indicative of an early chaotic phase of the evolution of
 the central regions of the Galaxy, as well. This is consistent with a scenario in which the bulge formed at
 high redshift through  processes such as mergers and violent disc instabilities, as recently confirmed
 by Tacchella et al. (2015). On the other hand, there is evidence that the Galactic bar is relatively recent
 with respect to the age of the bulge clusters. Cole \& Weinberg (2002), for example,  concluded that the
 bar must be younger than 6 Gyr. In this case the clusters formed before the development of the bar instability.
 An example could be NGC 6522, that is a very old cluster, $\sim13$ Gyr (Kerber et al. 2018) 
but has a bar/boxy-shape orbit (in the rotating frame), meaning that it was formed in a very early stage of the Galaxy, 
before bar formation, and it was trapped by the bar later on.



\begin{acknowledgements}
 We acknowledge the anonymous referee for enlightening comments that greatly improved this paper. APV acknowledges FAPESP for the postdoctoral fellowship 2017/15893-1.
 LR acknowledges a CRS scholarship from Swinburne University of Technology. BB  and  EB acknowledge  grants from  the
Brazilian  agencies  CNPq and  Fapesp.   SO  acknowledges the  Italian
Ministero dell'Universit\`a e della Ricerca Scientifica e Tecnologica, 
the Istituto Nazionale di Astrofisica (INAF) and the financial support of the University of Padova.
\end{acknowledgements}



\label{lastpage}

\begin{thebibliography}{}

\bibitem[Aguilar et al.(1988)]{1988ApJ...335..720A} 
Aguilar, L., Hut, P., \& Ostriker, J.~P.\ 1988, ApJ, 335, 720 

\bibitem[\protect\citeauthoryear{Allen et al.}{2006}]{2006ApJ...652.1150A} 
Allen, C., Moreno, E., \& Pichardo, B.\ 2006, ApJ, 652, 1150 

\bibitem[\protect\citeauthoryear{Allen et al.}{2008}]{2008ApJ...674..237A} 
Allen, C., Moreno, E., \& Pichardo, B.\ 2008, ApJ, 674, 237 

\bibitem[\protect\citeauthoryear{Barbuy et al.}{2018}]{Barbuy2018}
Barbuy, B., Chiappini, C., \& Gerhard, O, 2018, ARA\&A, in press

\bibitem[\protect\citeauthoryear{Benjamin et al.}{2005)}]{2005ApJ...630L.149B} 
Benjamin, R.~A., Churchwell, E., Babler, B.~L., et al.\ 2005, ApJL, 630, L149 

\bibitem[Berentzen \& Athanassoula(2012)]{2012MNRAS.419.3244B} 
Berentzen, I., \& Athanassoula, E.\ 2012, MNRAS, 419, 3244 

\bibitem[\protect\citeauthoryear{Bica, Ortolani, \& Barbuy}{2016}]{2016PASA...33...28B} 
Bica E., Ortolani S., Barbuy B., 2016, PASA, 33, e028 

\bibitem[Binney et al.(1991)]{1991MNRAS.252..210B} 
Binney, J., Gerhard, O.~E., Stark, A.~A., Bally, J., \& Uchida, K.~I.\ 1991, MNRAS, 252, 210 

\bibitem[Bissantz et al.(2003)]{2003MNRAS.340..949B} 
Bissantz, N., Englmaier, P., \& Gerhard, O.\ 2003, MNRAS, 340, 949 

\bibitem[\protect\citeauthoryear{Bland-Hawthorn \& Gerhard}{2016}]{2016ARA&A..54..529B} 
Bland-Hawthorn, J., Gerhard, O. 2016, ARA\&A, 54, 529 

\bibitem[Bobylev \& Bajkova(2017)]{2017ARep...61..551B} 
Bobylev, V.~V., \& Bajkova, A.~T.\ 2017, Astronomy Reports, 61, 551 

\bibitem[\protect\citeauthoryear{Casetti-Dinescu et al.}{2007}]{2007AJ....134..195C} 
Casetti-Dinescu, D.~I., Girard, T.~M., Herrera, D., et al.\ 2007, AJ, 134, 
195 

\bibitem[\protect\citeauthoryear{Casetti-Dinescu et al.}{2013}]{2013AJ....146...33C} 
Casetti-Dinescu, D.~I., Girard, T.~M., J{\'{\i}}lkov{\'a}, L., et al.\ 2013, AJ, 146, 33 

\bibitem[Chemin et al.(2015)]{2015A&A...578A..14C} 
Chemin, L., Renaud, F., \& Soubiran, C.\ 2015, A\&A, 578, A14 

\bibitem[\protect\citeauthoryear{Cole \& Weinberg}{2002}]{2002ApJ...574L..43C} 
Cole, A.~A., \& Weinberg, M.~D.\ 2002, ApJL, 574, L43 

\bibitem[\protect\citeauthoryear{Dinescu et al.}{2001}]{2001AJ....122.1916D} 
Dinescu, D.~I., Majewski, S.~R., Girard, T.~M., \& Cudworth, K.~M.\ 2001, AJ, 122, 1916 


\bibitem[Freudenreich(1998)]{1998ApJ...492..495F} 
Freudenreich, H.~T.\ 1998, ApJ, 492, 495 

\bibitem[\protect\citeauthoryear{Flynn et al.}{1996}]{1996MNRAS.281.1027F} 
Flynn, C., Sommer-Larsen, J., \& Christensen, P.~R.\ 1996, MNRAS, 281, 1027 

\bibitem[\protect\citeauthoryear{Gardner \& Flynn}{2010}]{2010MNRAS.405..545G} 
Gardner, E., \& Flynn, C.\ 2010, MNRAS, 405, 545 

\bibitem[Gnedin \& Ostriker(1997)]{1997ApJ...474..223G} 
Gnedin, O.~Y., \& Ostriker, J.~P.\ 1997, ApJ, 474, 223 

\bibitem[Gnedin et al.(1999)]{1999ApJ...522..935G} 
Gnedin, O.~Y., Lee, H.~M., \& Ostriker, J.~P.\ 1999, ApJ, 522, 935 

\bibitem[\protect\citeauthoryear{Hammersley et al.}{2000}]{2000MNRAS.317L..45H} 
Hammersley, P.~L., Garz{\'o}n, F., Mahoney, T.~J., L{\'o}pez-Corredoira, M., \& Torres, M.~A.~P.\ 2000, MNRAS, 317, L45 

\bibitem[Harris(2010)]{2010arXiv1012.3224H} 
Harris, W.~E.\ 2010, arXiv:1012.3224 

\bibitem[\protect\citeauthoryear{Irrgang et al.}{2013}]{2013A&A...549A.137I} 
Irrgang, A., Wilcox, B., Tucker, E., \& Schiefelbein, L.\ 2013, A\&A, 549, A137 

\bibitem[\protect\citeauthoryear{J{\'{\i}}lkov{\'a} et al.}{2012}]{2012A&A...541A..64J} 
J{\'{\i}}lkov{\'a}, L., Carraro, G., Jungwiert, B., \& Minchev, I.\ 2012, A\&A, 541, A64 

\bibitem[\protect\citeauthoryear{Kent et al.}{1991}]{1991ApJ...378..131K} 
Kent, S.~M., Dame, T.~M., \& Fazio, G.\ 1991, ApJ, 378, 131 

\bibitem[Kerber et al.(2018)]{2018ApJ...853...15K} 
Kerber, L.~O., Nardiello, D., Ortolani, S., et al.\ 2018, ApJ, 853, 15 

\bibitem[\protect\citeauthoryear{Long \& Murali}{1992}]{1992ApJ...397...44L} 
Long, K., \& Murali, C.\ 1992, ApJ, 397, 44 

\bibitem[Martinez-Medina et al.(2018)]{2018MNRAS.474...32M} 
Martinez-Medina, L.~A., Gieles, M., Pichardo, B., \& Peimbert, A.\ 2018, MNRAS, 474, 32 

\bibitem[Minniti et al. (2017a)]{Minniti2017a}
Minniti, D., Geisler, D., Alonso-Garcia, J. et al. 2017a, ApJ, 849, L24

\bibitem[Minniti et al. (2017b)]{Minniti2017b}
Minniti, D.; Alonso-Garc�a, J.; Pullen, J. 2017b, RNAAS, 1, 54

\bibitem[\protect\citeauthoryear{Miyamoto \& Nagai}{1975}]{1975PASJ...27..533M} 
Miyamoto, M., \& Nagai, R.\ 1975, PASJ, 27, 533 

\bibitem[\protect\citeauthoryear{Moreno et al.}{2014}]{2014ApJ...793..110M} 
Moreno, E., Pichardo, B., \& Vel{\'a}zquez, H.\ 2014, ApJ, 793, 110 

\bibitem[\protect\citeauthoryear{Navarro et al.}{1997}]{1997ApJ...490..493N} 
Navarro, J.~F., Frenk, C.~S., \& White, S.~D.~M.\ 1997, ApJ, 490, 493 

\bibitem[\protect\citeauthoryear{Ortolani et al.}{2011}]{2011ApJ...737...31O} 
Ortolani S., Barbuy B., Momany Y., Saviane I., Bica E., Jilkova L., Salerno G.~M., Jungwiert B., 2011, ApJ, 737, 31 

\bibitem[\protect\citeauthoryear{Pichardo et al.}{2004}]{2004ApJ...609..144P} 
Pichardo, B., Martos, M., \& Moreno, E.\ 2004, ApJ, 609, 144 

\bibitem[\protect\citeauthoryear{Pfenniger}{1984}]{1984A&A...134..373P} 
Pfenniger, D.\ 1984, A\&A, 134, 373

\bibitem[Portail et al.(2015)]{2015MNRAS.448..713P} 
Portail, M., Wegg, C., Gerhard, O., \& Martinez-Valpuesta, I.\ 2015, MNRAS, 448, 713 

 \bibitem[Portail et al.(2017)]{2017MNRAS.465.1621P} 
 Portail, M., Gerhard, O., Wegg, C., \& Ness, M.\ 2017, MNRAS, 465, 1621 

\bibitem[\protect\citeauthoryear{Prugniel \& Simien}{1997}]{1997A&A...321..111P} 
Prugniel, P., \& Simien, F.\ 1997, A\&A, 321, 111 

\bibitem[Rattenbury et al.(2007)]{2007MNRAS.378.1064R} 
Rattenbury, N.~J., Mao, S., Sumi, T., \& Smith, M.~C.\ 2007, MNRAS, 378, 1064 

\bibitem[\protect\citeauthoryear{Rossi}{2015a}]{2015ascl.soft01002R} 
Rossi, L.~J.\ 2015a, Astrophysics Source Code Library, 1501.002 

\bibitem[\protect\citeauthoryear{Rossi}{2015b}]{2015A&C....12...11R} 
Rossi, L.~J.\ 2015b, Astronomy and Computing, 12, 11 

\bibitem[\protect\citeauthoryear{Rossi et al.}{2015}]{2015MNRAS.450.3270R} 
Rossi L.~J., Ortolani S., Barbuy B., Bica E., Bonfanti A., 2015, MNRAS, 450, 3270 

\bibitem[Sch{\"o}nrich et al.(2010)]{2010MNRAS.403.1829S} 
Sch{\"o}nrich, R., Binney, J., \& Dehnen, W.\ 2010, MNRAS, 403, 1829 

\bibitem[\protect\citeauthoryear{Smith et al.}{2015}]{2015MNRAS.448.2934S} 
Smith, R., Flynn, C., Candlish, G.~N., Fellhauer, M., \& Gibson, B.~K.\ 2015, MNRAS, 448, 2934 

\bibitem[\protect\citeauthoryear{Sohn et al.}{2015}]{2015ApJ...803...56S} 
Sohn S.~T., van der Marel R.~P., Carlin J.~L., Majewski S.~R., Kallivayalil N., Law D.~R., Anderson J., Siegel M.~H., 2015, ApJ, 803, 56 

\bibitem[\protect\citeauthoryear{Tacchella et al.}{2015}]{2015Sci...348..314T} 
Tacchella S., et al., 2015, Sci, 348, 314 

\bibitem[Terndrup et al.(1998)]{1998AJ....115.1476T} 
Terndrup, D.~M., Popowski, P., Gould, A., Rich, R.~M., \& Sadler, E.~M.\ 1998, AJ, 115, 1476 

\bibitem[\protect\citeauthoryear{Terzi{\'c} \& Graham}{2005}]{2005MNRAS.362..197T} 
Terzi{\'c}, B., \& Graham, A.~W.\ 2005, MNRAS, 362, 197 

\bibitem[Wegg \& Gerhard(2013)]{2013MNRAS.435.1874W} 
Wegg, C., \& Gerhard, O.\ 2013, MNRAS, 435, 1874 

\bibitem[\protect\citeauthoryear{Weiner \& Sellwood}{1999}]{1999ApJ...524..112W} 
Weiner, B.~J., \& Sellwood, J.~A.\ 1999, ApJ, 524, 112 


\end{thebibliography}
\end{document}